\documentclass[pra]{revtex4}
\usepackage{amssymb}
\usepackage{amsmath}

\newcommand{\vc}[1]{\boldsymbol{#1}}

\begin{document}

\title{Hamiltonian, Geometric Momentum and Force Operators for a Spin Zero Particle on a Curve: Physical Approach }

\author{M. S. Shikakhwa}
\affiliation{Department of Basic Sciences, TED University\\ Ziya Gökalp Caddesi No:48, 06420, Ankara, Türkiye}

\author{N.Chair}
\affiliation{Department of Physics,University of Jordan,\\Queen Rania Street,\\
Amman, Jordan}

\begin{abstract}
The  Hamiltonian for a spin zero particle that is confined to a curve embedded in the 3D space is 
constructed by squeezing the coordinates spanning a tube normal to the curve onto the curve assuming  strong normal forces. 
We follow the new approach that we applied to confine a particle to a surface, in that we start with an expression for the 3D momentum operators
whose components along and  normal to the curve directions are separately Hermitian. The
kinetic energy operator expressed in terms of the momentum operator in the normal direction is then a Hermitian operator in this case. When this
operator is dropped and the thickness of the tube surrounding the curve is set to zero, one automatically gets the Hermitian
curve Hamiltonian that contains the geometric potential term as expected. It is demonstrated that the origin of this potential lies in the ordering or symmetrization of the original 3D momentum operators in order to render them Hermitian. The Hermitian momentum operator for the particle as it is 
confined to the curve  is also constructed and is seen to be similar to what is known as the geometric momentum of a particle  confined to a surface in that it has a term proportional to the curvature that is along the normal to the curve. The force operator of the particle on the curve is also derived, and is shown to reduce, for a curve with a constant curvature and torsion, to a -apparently-  single component normal to the curve that is a symmetrization of the classical expression plus a quantum term. All the above quantities are then derived for the specific case of a particle confined to a cylindrical helix embedded in 3D space. 
\end{abstract}

\maketitle
\section{Introduction}
The research on the non-relativistic quantum mechanics on general 2D surfaces and 1D curves, despite being intrinsically interesting, has received a substantial boost in the last decades due to the fast technological advances in the fabrication of nano-scale structures, like nano-spheres, nano-tubes...etc. A - by now- popular approach to the problem of constructing the Hamiltonian is such systems was first developed by Jensen and Koppe \cite{Koppe} for confining a particle to a spherical surface, then was generalized to confinement to an arbitrary 2D surface embedded in 3D space by da Costa \cite{Costa} and we refer to it now on by the CJK approach ( also known in the literature as Thin Layer Quantization (TLQ)). da Costa in \cite{Costa} applied the approach to confine a spin zero particle to an arbitrary curve, too. Later works \cite{Ortix,Wang,Magarill} applied the same approach to construct the Hamiltonian of a spin one-half particle in the presence of a variety of interactions and couplings confined to a 1D curve embedded in 3D Euclidean space. The essential idea is to start with the Schrödinger equation for a particle in a 3D layer ( a 3D tube for confinement to a 1D curve)  surrounding a surface (a curve). A very strong confining potential depending only on the coordinate(s) normal to, but not those along the surface (curve) is assumed to pin the particle to the surface ( curve). Eventually the two (one) coordinate(s) normal to the surface (curve) are (is) set to zero and the resulting Schrödinger equation is separated into two equations for the normal and surface (curve) degrees of freedom. The key point in the CJK approach is a transformation of the wavefunction that isolates a finite term that would otherwise be mistakenly dropped from the
Hamiltonian once the thickness of the layer ( tube) is set to zero. The authors of \cite{Koppe} call this the dangerous term. This finite term gives in the zero thickness limit what is known as  the geometrical potential or the geometrical kinetic energy contribution to the Hamiltonian. This potential appears when pinning a particle to both a surface and a curve and is essential to render the Hamiltonian Hermitian. Recently, we introduced \cite {shikakhwa} a twist to the CJK approach to constraint a particle to a surface, which renders it more intuitive and transparent, we believe. Our idea is to note that while, intuitively, dropping the derivatives with respect to the coordinates normal to the surface  when pinning the particle to it, by setting the coordinates normal to the surface to zero should work well, we need to drop Hermitian quantities to end up with a Hermitian Hamiltonian on the surface. We achieve this by working from the onset with momentum operators along and normal to the surface that are separately Hermitian. This way, we drop, once the particle is confined to the reduced manifold, the Hermitian normal momentum operator and are left with a Hermitian Hamiltonian with the geometric potential appearing naturally. We have applied our approach to confine a spin zero particle to an arbitrary surface embedded in 3D space spanned by general curvilinear coordinates in \cite{shikakhwa} and to confine a spin one-half particle with various interactions to a 2D surface embedded in a 3D space spanned by orthogonal curvilinear coordinates in \cite{shikakhwa and chair1,shikakhwa and chair2}.\\
In the present work, we show that our approach can also be applied to the problem of confining a spin zero particle to an arbitrary space curve embedded in 3D Euclidean space spanned by general curvilinear coordinates. We construct the Hermitian 3D momentum operators along and normal to the curve that we need to start with (section 1)and derive the Hermitian Hamiltonian upon pinning the particle to the curve in section 2. In section 3, we derive the expression of the Hermitian momentum operator of the particle once it is pinned to the curve and show that it is exactly similar to the so called geometrical momentum on a surface \cite{shikakhwa symmetric,{Liu2}}. In section 4, we derive the force operator for the particle on the curve and in section 5 we derive all the above quantities for the specific example of a particle confined to a cylindrical helix embedded in 3D. We sum up and give our conclusions in section 6.  
   
\section{The 3D Hermitian momentum operator }
To confine a particle in 3D space to an arbitrary space curve embedded in this space, we start with the coordinate system -adopted also in \cite{Ortix, Wang}- where a vector $\boldsymbol{a}(q_1)$ centered at the origin of the coordinates traces the curve as the coordinate $q_1$ varies along the curve. At each point on the curve, i.e. for each value of $q_1$, the coordinates $q_{2}$ and $q_{3}$ vary over the  normal $\hat{n}(q_1)$ and binormal $\hat{b}(q_1)$ unit vectors, respectively, defining a curve-centered 3D coordinate system. The position vector $\boldsymbol{R}\left(q_1, q_{2}, q_{3}\right)$ of any point in space in the neighbourhood of the curve is then given as: 
\begin{equation}\label{coordinate}
\boldsymbol{R}\left(q_1, q_{2}, q_{3}\right)=\boldsymbol{a}(q_1)+q_{2} \hat{n}(q_1)+q_{3} \hat{b}(q_1)
\end{equation}
 $\hat{t}(q_1)=\partial{\boldsymbol{a}}/\partial{q_1}$ is a unit vector tangent to the curve forming with $\hat{n}(q_1)$ and $\hat{b}(q_1)$ a set of three orthogonal unit vectors defining the so called Fresnel frame. These  three unit vectors are related by the well-known Frenet-Serret equations:
\begin{equation}\label{FS}
\left(\begin{array}{c}
\partial_{1} \hat{t}(q_1) \\
\partial_{1} \hat{n}(q_1) \\
\partial_{1} \hat{b}(q_1)
\end{array}\right)=\left(\begin{array}{ccc}
0 & \kappa(q_1) & 0 \\
-\kappa(q_1) & 0 & \tau(q_1) \\
0 & -\tau(q_1) & 0
\end{array}\right)\left(\begin{array}{c}
\hat{t}(q_1) \\
\hat{n}(q_1) \\
\hat{b}(q_1)
\end{array}\right)
\end{equation}

with $\kappa(q_1)$ and $\tau(q_1)$ denoting the curvature and torsion of the space curve, respectively.
The  vectors  tangent to the coordinates $\boldsymbol{u}_{i} \equiv \frac{\partial \boldsymbol{R}}{\partial q^{i}}=\partial_{i} \boldsymbol{R}$ are defined as usual, and for our coordinate system they read: 
\begin{equation}\label{u1}
\begin{aligned}
\boldsymbol{u}_{1} \equiv \partial_{1} \boldsymbol{R}=&\partial_{1} \boldsymbol{a}+  q_{2} \partial_{1} \hat{n}(q_1)+q_{3} \partial_{1} \hat{b}(q_1)  \\
 =&\hat{t}(q_1)+  q_{2} \partial_{1} \hat{n}(q_1)+q_{3} \partial_{1} \hat{b}(q_1)\\
 =&\hat{t}(1-q_2\kappa)-q_3\tau\hat{n}+q_2\tau\hat{b}
 \end{aligned}
\end{equation}
and, 
\begin{equation}\label{u2 and u3}
\begin{aligned}
 \boldsymbol{u}_{2} &\equiv& \partial_{2} \boldsymbol{R}=\hat{n}(q_1)\\
\boldsymbol{u}_{3} &\equiv& \partial_{3} \boldsymbol{R}=\hat{b}(q_1)
\end{aligned}
\end{equation}
 where Eq.(\ref{FS}) was used to express $\boldsymbol{u}_{1}$ in terms of $\kappa(q_1)$ and $\tau(q_1)$ in Eq.(\ref{u1}). 
 The metric tensor $G_{i j}=\boldsymbol{u}_i\cdot \boldsymbol{u}_j$ following from the above coordinate system then reads: 
\begin{equation}\label{Gij}
G_{i j}=\left(\begin{array}{ccc}
\left(1-\kappa q_2\right)^2+\tau^2\left(q_2^2+q_3^2\right) & -\tau q_3 & \tau q_2 \\
-\tau q_3 & 1 & 0 \\
\tau q_2 & 0 & 1
\end{array}\right)
\end{equation}

The determinant $G$ of this metric tensor is then $ G=det G_{i j}=\left(1-\kappa q_{2}\right)^{2}$. The inverse metric $G^{ij}=\frac{K^{ij}}{G}$ with $K^{ij}$ being the cofactor of $G_{ij}$  is then calculated to read:
\begin{equation}
G^{i j}=\frac{1}{\left(1-\kappa q_{2}\right)^{2}}\left(\begin{array}{ccc}
\left(1-\kappa q_2\right)^2+\tau^2\left(q_2^2+q_3^2\right) & -\tau^2 {q_3}^2 & -\tau^2 {q_2}^2 \\
-\tau^2 {q_3}^2& \left(1-\kappa q_2\right)^2+ \tau^2 {q_3}^2& 0 \\
-\tau^2 {q_2}^2& 0 & \left(1-\kappa q_2\right)^2+ \tau^2 {q_2}^2
\end{array}\right)
\end{equation}
We now construct the Hermitian momentum operators along and perpendicular to the curve. The Hermicity of the 3D momentum operator $\boldsymbol{p}=-i \hbar \nabla$ should evidently be preserved when it is expressed in general curvilinear coordinates where it reads $\boldsymbol{p}=-i \hbar \nabla=-i \hbar \boldsymbol{u}^{i} \partial_{i}$ (recall that $\boldsymbol{u}_{i} \equiv \partial_{i} \boldsymbol{R}$ and so $\boldsymbol{u}^{i}=G^{ij}\boldsymbol{u}^{j}$). Checking, we find:

$$
\langle\Psi \mid \boldsymbol{p} \Psi\rangle=\langle\boldsymbol{p} \Psi \mid \Psi\rangle+\left\langle\Psi \mid \frac{-i \hbar}{\sqrt{G}} \partial_{i}\left(\sqrt{G} \boldsymbol{u}^{i}\right) \Psi\right\rangle
$$

where integration is over all space with the measure $\sqrt{G} d^{3} u$ and the wavefunction was assumed to satisfy boundary conditions that allows the surface term to be dropped. Hermicity of $\boldsymbol{p}$ demands the vanishing of the second term on the r.h.s, i.e.

\begin{equation}\label{gamma}
\frac{1}{\sqrt{G}} \partial_{i}\left(\sqrt{G} \boldsymbol{u}^{i}\right)=\frac{1}{\sqrt{G}} \partial_{1}\left(\sqrt{G} {\boldsymbol{u}}^{1}\right)+\frac{1}{\sqrt{G}} \partial_{a}\left(\sqrt{G} \boldsymbol{u}^{a}\right)=0
\end{equation}

where the tangent vectors $\boldsymbol{u}_{a}, a=1,2$ are, respectively, just $\hat{n}(q_1)$ and $\hat{b}(q_1)$ defined in Eq.(\ref{u2 and u3}). The relation $\frac{1}{\sqrt{G}} \partial_{i}\left(\sqrt{G} \boldsymbol{u}^{i}\right)=0$ is in fact an identity as was discussed in \cite{shikakhwa}.The 3D momentum operator $\boldsymbol{p}$ when expressed  as a sum of two momentum operators $\boldsymbol{p}_{1}$ and $\boldsymbol{p}_{\perp}$  along and normal to the curve, respectively, reads:

\begin{equation}\label{p}
\begin{aligned}
\boldsymbol{p} & =\boldsymbol{p}_{1}+\boldsymbol{p}_{\perp}\\
& =-i \hbar {\boldsymbol{u}}^{1} \partial_{1}+-i \hbar \boldsymbol{u}^{a} \partial_{a}
\end{aligned}
\end{equation}

It is straightforward to check that neither the momentum operator along  nor that normal to the curve are Hermitian in 3D space as they stand; only their sum is. Now adding (half) of the zero-valued expression in Eq. (\ref{gamma}) multiplied by $-i \hbar$  split among $\boldsymbol{p}_{1}$ and $\boldsymbol{p}_{\perp}$ to the above $\boldsymbol{p}$, we have:

\begin{equation}\label{pH}
\boldsymbol{p}=\boldsymbol{p}_{1}+\boldsymbol{p}_{\perp}=\boldsymbol{p}_{1H}+\boldsymbol{p}_{\perp H}
\end{equation}

with the operators $\boldsymbol{p}_{1H}$ and $\boldsymbol{p}_{\perp H}$ being now:
\begin{equation}\label{p1H}
\begin{aligned}
\boldsymbol{p}_{1H} & =\boldsymbol{p}_{1}-\frac{i \hbar}{2 \sqrt{G}} \partial_{1}\left(\sqrt{G} {\boldsymbol{u}}^{1}\right) \\
& =-i \hbar \left({\boldsymbol{u}}^{1}\partial_{1}+\mathbf{\Gamma}\right)
\end{aligned}
\end{equation}
and,
\begin{equation}\label{p perp H}
\begin{aligned}
\boldsymbol{p}_{\perp H}& =\boldsymbol{p}_{\perp}-\frac{i \hbar}{2 \sqrt{G}} \partial_{a}\left(\sqrt{G} \boldsymbol{u}^{a}\right)=\boldsymbol{p}_{\perp}+\frac{i \hbar}{2 \sqrt{G}} \partial_{1}\left(\sqrt{G} {\boldsymbol{u}}^{1}\right) \\
& =-i \hbar\left(\boldsymbol{u}^{a} \partial_{a}- \mathbf{\Gamma}\right)
\end{aligned}
\end{equation}

and we have defined $\mathbf{\Gamma}$ as:
\begin{equation}\label{gamma defined}
\mathbf{\Gamma}=\frac{1}{2 \sqrt{G}} \partial_{1}(\sqrt{G}{\boldsymbol{u}}^{1})
\end{equation}

 The newly defined  momentum operator along the curve $\boldsymbol{p}_{1H}$ and the momentum operator normal to the curve $\boldsymbol{p}_{\perp H}$ can be readily checked to be Hermitian over the 3D space. Therefore, by adding a zero-valued quantity to the full 3D Hermitian momentum operator, we managed to express it as the sum of two Hermitian  momentum operators; along and normal to the curve. This a key step for the following analysis.
\section{The Hermitian Hamiltonian on the Curve}

In this section, we will construct the Hamiltonian for a spin zero particle confined to the space curve and is otherwise free. The approach \cite{shikakhwa} is based on the intuitive argument that if one starts from the full Hamiltonian in the 3D space spanned by the coordinate system given in Eq. (\ref{coordinate}) and then confine the particle to the curve by introducing a strong confining potential (force) along the direction normal to the curve; then, the excitation along this direction will need an infinite energy and so the dynamics is essentially along the curve. This amounts to freezing the normal degree of freedom and dropping it from the Hamiltonian which is achieved by setting $q_2,q_3$ to zero and dropping the normal momentum operator from the kinetic energy operator. The critical point of the present approach is to use in the Hamiltonian  momentum (and so kinetic energy) operators along and normal to the curve that are \textit{separately} Hermitian, so that  as one drops the Hermitian kinetic energy operator ( Hermitian momentum normal to the curve)  one is left with a Hermitian  Hamiltonian on the curve. While this intuition might seem obvious and trivial, blindly dropping the normal degrees of freedom without observing for Hermicity has led to the reporting of non-Hermitian surface Hamiltonians in the literature \cite{Aronov}. Therefore, the essential starting point is the expression of the 3D momentum operator as a sum of the two Hermitian surface and normal momentum operators as in Eqs. (\ref{p1H}) and (\ref{p perp H}). The free particle Hamiltonian in 3D with the momentum operator given by these two equations reads:
\begin{equation}\label{H expanded}
\begin{aligned}
H_{3D}=\frac{p^{2}}{2 m}=&\frac{1}{2 m}\left(p^{2}_{1H}+p_{\perp H}^{2}+\mathbf{p}_{1 H} \cdot \boldsymbol{p}_{\perp H}+\boldsymbol{p}_{\perp H} \cdot \boldsymbol{p}_{1 H}\right)\\
=&\frac{p_{\perp H}^2}{2 m}-\frac{\hbar^2}{2 m} \left(\mathbf{u}^1 \cdot\left(\partial_1 \mathbf{u}^a\right) \partial_a+\mathbf{u}^a \cdot\left(\partial_a \mathbf{u}^1\right) \partial_1\right. \\
& +\mathbf{u}^1 \cdot\left(\partial_1 \mathbf{u}^1\right) \partial_1+2 \mathbf{u}^1 \cdot \mathbf{u}^a \partial_1 \partial_a \\
& +\mathbf{u}^1 \cdot \mathbf{u}^1 \partial_1^2+2 \mathbf{u}^a \cdot \mathbf{\Gamma}\partial_a+\mathbf{u}^a \cdot \partial_a \mathbf{\Gamma} \\
& -\mathbf{\Gamma} \cdot \mathbf{\Gamma})
\end{aligned}
\end{equation}
In the limit $q_2, q_3 \rightarrow 0$, we note that
\begin{equation}\label{limit 1}
\begin{aligned}
& G_{i j}, G^{i j} \longrightarrow \delta_{i j}, \delta^{i j} \\
& G \longrightarrow 1\\
& \mathbf{u}^2=G^{2 j} \mathbf{u}_j \longrightarrow \hat{n} \\
& \mathbf{u}^3=G^{3 j} \mathbf{u}_j \rightarrow \hat{b} \\
& \mathbf{u}^1=G^{1 j} \mathbf{u}_j \rightarrow \hat{t}
\end{aligned}
\end{equation}
leading in this limit to:
\begin{equation}\label{limit 2}
\begin{aligned}
 &\mathbf{u}^1 \cdot\left(\partial_1 \mathbf{u}^a\right) \partial_a \longrightarrow-\kappa \partial_2 \\
 &2 \mathbf{u}^a \cdot \mathbf{\Gamma} \partial_a\longrightarrow\kappa \partial_2\\
 &\mathbf{u}^1 \cdot \mathbf{u}^1 \partial_1^2 \longrightarrow \partial_1^2 \\
 &\mathbf{u}^a \cdot \partial_a \mathbf{\Gamma} \longrightarrow \kappa^2 / 2 \\
 &\mathbf{\Gamma} \cdot \mathbf{\Gamma}\longrightarrow \kappa^2 / 4 \\
\end{aligned}
\end{equation}
and all of $\mathbf{u}^a \cdot\left( \partial_a \mathbf{u}^1\right) \partial_1 ; \mathbf{u}^1 \cdot\left(\partial_1 \mathbf{u}^1\right) \partial_1 ; 2 \mathbf{u}^1 \cdot \mathbf{u}^a \partial_1 \partial_a \longrightarrow 0$. Along with taking the limit $q_2, q_3 \rightarrow 0$ we also drop the Hermitian kinetic energy operator representing the degrees of freedom normal to curve, i.e. $\frac{p_{\perp H}^2}{2 m}$ from the Hamiltonian. This immediately gives the same Hamiltonian as in TLQ with the geometric potential $-\frac{\hbar^2}{2 m}\left(\frac{\kappa^2}{4}\right)$ appearing automatically:
\begin{equation}\label{H curve}
H=-\frac{\hbar^2}{2 m}\left(\partial_1^2+\frac{\kappa^2}{4}\right) \\
\end{equation}
%This reduces the Hamiltonian to:
%$$
%H=\frac{-\hbar^2}{2 m}\left(\partial_1^2+\frac{\kappa^2}{4}\right)+\left.\frac{p_{\perp H}^2}{2 m}\right|_{q_2,q_3 \rightarrow 0}
%$$
%Setting $P_{\perp H}^2 \rightarrow 0$, we end up with
\section{The geometric Momentum on the Curve}
In Eq.(\ref{p1H}) we have given the Hermitian momentum vector along the curve in 3D space, i.e. before pinning the particle to the curve as: 
\begin{equation}
\mathbf{p}_{1 H}=-i \hbar\left(\mathbf{u}^1 \partial_1+\frac{1}{2 \sqrt{G}} \partial_1 (\sqrt{G} \mathbf{u}^1)\right)
\end{equation}
Here, we find the form to which this operator reduces in the limit $q_2, q_3 \rightarrow 0$ and check for its Hermicity in the reduced 1D space. In the limit $q_2, q_3 \rightarrow 0, \frac{1}{2 \sqrt{G}} \partial_1 \sqrt{G} \mathbf{u}^1=\mathbf{{\Gamma}} \rightarrow \frac{\kappa}{2} \hat{n}$ reducing the momentum along the curve to the expression:
\begin{equation}\label{P curve Hermitian}
\mathbf{p}_{1H} \longrightarrow\mathbf{p}^c_{1H}=-i \hbar\left(\hat{t} \partial_1+\frac{\hat{n} \kappa}{2}\right)
\end{equation}

This curve momentum operator $\mathbf{p}^c_{1H}$is Hermitian on the reduced 1-D space, i.e
\begin{equation}
\int d q_1 \psi^* \mathbf{p}^c_{1 H} \psi=\int d q_1\left(\mathbf{p}^c_{1H} \psi\right)^* \psi
\end{equation}
The $\frac{\hat{n} \kappa}{2}$ term is essential to establish Hermicity; $-i \hbar \hat{t} \partial_1$ is not Hermitian by itself. The Hermitian momentum on a curved surface embedded in 3D space has a form similar to this curve  momentum, Eq.(\ref{P curve Hermitian}), and is sometimes called the geometric momentum in the literature. It was derived using Dirac quantization in \cite{Liu2} and using an approach similar to the current one in \cite{shikakhwa, shikakhwa symmetric}. This is - to the best of our knowledge- the first time it is reported for a curve. 
The fact that the normal vector $\hat{n}$ appears in the momentum operator on the curve, Eq.(\ref{P curve Hermitian}), does not mean that it has a normal component. In fact one can easily check that  $\hat{n}\cdot\mathbf{p}^c_{1H}+\mathbf{p}^c_{1H}\cdot\hat{n}=0$ indicating that it is purely tangential. In fact, just as is the case with the geometric momentum on a surface \cite{shikakhwa}, this curve geometric momentum operator can be expressed in a symmetric form that makes it manifestly tangential:
\begin{equation}\label{p curve symmetric} 
 \mathbf{p}^c_{1H} =-i \frac{\hbar}{2}\left(\hat{t} \partial_1+\partial_1\hat{t}\right)
\end{equation}
where in the second term of the above equation, the $\partial_1$ is to be understood as acting on everything to its right. It is interesting to note that the curve geometric momentum operator, $\mathbf{p}^c_{1H}$, is the kinematical momentum operator that we get when taking the time derivative of $\boldsymbol{a}(q_1)$, the position operator of the particle on the curve ( see Eq.(\ref{coordinate})). Indeed:
$$
m \frac{d \boldsymbol{a}}{d t}=\frac{[\boldsymbol{a},H]}{i \hbar}=-i \hbar\left(\hat{t} \partial_1+\frac{\hat{n} \kappa}{2}\right)=\mathbf{p}^c_{1H}
$$
Closing this section we would like to note it is not only the Hermitian momentum on the curve that can be written in a symmetric form as in Eq.(\ref{p curve symmetric}) above. In fact the Hermitian momentum operators in 3D curvilinear coordinate, Eq.(\ref{p1H}) and Eq.(\ref{p perp H}), can also be symmetrized as was discussed in \cite{shikakhwa} to read :
\begin{equation}\label{p1 symmetrized}
\begin{aligned}
 \boldsymbol{p}_{1H} =&\boldsymbol{p}_{1}-\frac{i \hbar}{2 \sqrt{G}} \partial_{1}\left(\sqrt{G} {\boldsymbol{u}}_{1}\right)=-i \hbar \left({\boldsymbol{u}}_{1}\partial_{1}+\mathbf{\Gamma}\right) \\
= &\frac{-i\hbar}{2\sqrt{G}}(\sqrt{G}\vc{u}^1\partial_1+\partial_1\sqrt{G}\vc{u}^1)=\frac{-i\hbar}{2\sqrt{G}}\{\sqrt{G}\vc{u}^1,\partial_1\}_{+}
\end{aligned}
\end{equation}
and,
\begin{equation}\label{p perp symmetrized}
\begin{aligned}
 \boldsymbol{p}_{\perp H}& =\boldsymbol{p}_{\perp}-\frac{i \hbar}{2 \sqrt{G}} \partial_{a}\left(\sqrt{G}\right)=-i \hbar\left(\boldsymbol{u}^{a} \partial_{a}- \mathbf{\Gamma}\right)  \\
= &\frac{-i\hbar}{2\sqrt{G}}(\sqrt{G}\vc{u}^a\partial_a+\partial_a\sqrt{G}\vc{u}^a)=\frac{-i\hbar}{2\sqrt{G}}\{\sqrt{G}\vc{u}^a,\partial_a\}_{+}
\end{aligned}
\end{equation}
with $a$ running over 2 and 3, and the bracket $\{.,.\}_{+}$ denotes anti-commutator. Now, we have seen in the derivations made in the Hamiltonian, Eq.(\ref{H expanded}), that it is the  $\mathbf{\Gamma}$ that gives rise to the appearance of the geometrical potential in the Hamiltonian. But this $\mathbf{\Gamma}$, which is added to the two momentum operators in Eqs.(\ref{p1H}) and (\ref{p perp H})  to render each Hermitian  is nothing but a recipe for symmetrizing the derivatives in the specific manner given in these equations in order to construct the 3D Hermitian operator $\boldsymbol{p}$. This means that the geometric potential has its origin in the symmetrization or ordering of the derivatives in the momentum operators. The same argument was made in the case of a spin-zero particle confined to a surface \cite{shikakhwa}.  
\section{The Force Operator}
The force operator on the particle on the curve is one more quantity that we can calculate. The Heisenberg equation for the curve geometric momentum gives:
\begin{equation}\label{F derived}
\begin{aligned}
& F=\frac{d \mathbf{p}^c_{1H}}{d t}=\frac{\left[\mathbf{p}^c_{1H}, H\right]}{i \hbar} \\
& =\frac{\hbar^2}{2 m}\left([\hat{t} \partial_1, \partial_1^2]+[\hat{t} \partial_1, \frac{\kappa^2}{4}]+\frac{1}{2}[\hat{n} \kappa, \partial_1^2]\right) \\
& =\frac{\hbar^2}{2 m}\left(\hat{t}(2 \kappa^2 \partial_1 +2\kappa \kappa^{\prime})+\hat{n}(-2 \kappa \partial_1^2-2 \kappa^{\prime} \partial_1 +\frac{\kappa^3}{2}+\frac{\tau^2 \kappa}{2}-\frac{\kappa^{''}}{2})\right. \\
&+\left.\hat{b}(-2 \tau \kappa \partial_1-\frac{\kappa \tau^{\prime}}{2} -\tau \kappa^{\prime})\right) \\
\end{aligned}
\end{equation}
The $\kappa\kappa^{'}$ - term along $\hat{t}$ has two origins: The standard $\left[\hat{t} \hat{\partial}_1, \frac{\kappa^2}{u}\right]$ term which is the gradient of the geometric potential, plus the $\left[ \hat{n} \kappa, \partial_1^2\right]$-term. The $\hat{t}2 \kappa^2 \partial_1$ originates from this last term, too. This means that, even if the curvature is constant so that it has no dependence on $q_1$, the force operator for a particle on a curve will always have a tangential component, in contrast to the classical case.  The torsion and its derivatives have their origin in the $\frac{1}{2}[\hat{n} \kappa, \partial_1^2]$-term. The above expression can be brought to a more symmetrical and transparent form:
\begin{equation}\label{F compact}
\begin{aligned}
& F=\frac{\hat{n}}{2} \kappa m v^2+m v^2 \kappa \frac{\hat{n}}{2}-\hat{n} \frac{\hbar^2 \kappa}{4 m}\left(2 \kappa^2+\tau^2-\frac{\kappa^{\prime \prime}}{\kappa}\right) \\
& -\frac{\hbar^2}{2 m}\left(\hat{t}(\kappa \kappa^{\prime})-\hat{b}(\frac{\tau \kappa^{\prime}}{2}+\kappa^{\prime} \tau)\right) \\
\end{aligned}
\end{equation}
where we have defined
$$
v^2=\frac{({\mathbf{p}^{c}_{1H}})^2}{m^2}=\frac{-\hbar^2}{m^2}\left(\partial_1^2-\frac{\kappa^2}{4}\right)
$$
Non -constant curvature and torsion give rise to a force with explicit components along the tangential and binormal directions. While the force along the normal direction in Eq.(\ref{F compact}) seems like a symmetrization of the classical expression plus a quantum term,  we need to keep in mind that $v$  contains derivative with respect to $q_1$ and that $\kappa=\kappa(q_1)$ and $\hat{n}=\hat{n}(q_1)$. In particular,  this symmetrization hides a tangential component of the force. For constant $\kappa$ and $\tau$  the force reduces to a simpler expression that is -again, apparently- a symmetrization of the classical one plus a quantum term:
\begin{equation}\label{F symmetrical}
F=\frac{\hat{n}}{2} \kappa m v^2+m v^2 \kappa \frac{\hat{n}}{2}-\hat{n} \frac{\hbar^2\kappa}{4 m}\left(2 \kappa^2+\tau^2\right)
\end{equation}
We  derived similar symmetrical expressions for the force on a particle confined to specific two-dimensional surfaces embedded in 3D space in \cite{shikakhwa symmetric}.

\section{Momentum, Hamiltonian and Force for a particle on a Helix}
A cylindrical helix characterized by radius $R$ and apex $C$ is given in cylindrical coordinates as  $r=R, \theta=\theta(s), z=C \theta(s)$ where $s=\sqrt{R^2+C^2} \theta$ is the arc  subtended by $\theta$. The unit vectors $\hat{t}$, $\hat{n}$ and $\hat{b}$ of the Fresnel coordinate system on the curve are given for the helix in terms of the cylindrical unit vectors as
\begin{equation}
\begin{aligned}
& \hat{t}=\frac{R \hat{\theta}+C \hat{z}} {\left(R^2+C^2\right)^{\frac{1}{2}}}=(\kappa \hat{\theta}+\tau \hat{z})\left(R^2+C^{2}\right)^{\frac{1}{2}} \\
& \hat{n}=-\hat{r} \\
& \hat{b}=\frac{R \hat{z}-C \hat{\theta}}{\left(R^2+C^2\right)^{\frac{1}{2}}}=(\kappa \hat{z}-\tau \hat{\theta})\left(R^2+C^2\right)^{\frac{1}{2}} \\
&
\end{aligned}
\end{equation}

so that
\begin{equation}
\partial_1=\left(R^2+C^2\right)^{-\frac{1}{2}}\partial_\theta=\partial_s
\end{equation}
In this coordinate system, we have $\kappa=\frac{R} { R^2+C^2}$ and $\tau=\frac{C}{R^2+C^2}$. It  is convenient to define the constant angle  $\alpha$ by $ \tan \alpha=\frac{\kappa}{\tau}=\frac{R}{C}$. For the helix, the geometric momentum, Eq.(\ref{P curve Hermitian}), and the Hamiltonian, Eq.(\ref{H curve}) are given, respectively, by:
\begin{equation}
\mathbf{p}_{hlx}  =-i\hbar\left((\kappa \hat{\theta}+\tau \hat{z}) \partial_\theta-\frac{\hat{r} \kappa}{2}\right) 
\end{equation}
\begin{equation}
 H_{hlx}  =\frac{-\hbar^2}{2m\left(R^2+C^2\right)}\left(\partial_\theta^2+\frac{\sin ^2 \alpha}{4}\right)
\end{equation}
where $\sin \alpha=\frac{R^2} { R^2+C^2}$. The force operator, with $\kappa$ and $\tau$ being constants for the helix can be readily written from Eq,(\ref{F symmetrical}), and reads:
\begin{equation}
\begin{aligned}
& F_{hlx}=\frac{\sin \alpha}{2}\left((-\hat{r}) \frac{m v^2}{R}+\frac{m v^2}{R}(-\hat{r})\right)+\frac{\hbar^2 R \hat{r}}{4m\left(R^2+c^2\right)^3}\left(2 \sin ^2 \alpha+\cos ^2 \alpha\right) \\
\end{aligned}
\end{equation}
with
\begin{equation}
 v^2=\frac{p_{hlx}^2}{m^2}=\frac{-\hbar^2}{m^2\left(R^2+c^2\right)}\left(\partial_\theta^2-\frac{\sin ^2 \alpha}{4}\right) \\
\end{equation}

\section{conclusion}
The Hermitian Hamiltonian of a spin zero particle on an arbitrary space curve embedded in Euclidean 3D space can be constructed  applying the following intuitive and physical procedure: Start with the 3D Hamiltonian of a particle in the vicinity of the curve formulated using the specific coordinate system, Eq.(\ref{coordinate}), with one coordinate $q_1$ along the curve and the other two, $q_2$ and $q_3$ normal to the curve. Then, re-express the Hamiltonian so that each of the momentum operators along and normal to the curve are separately Hermitian, which can be done by adding to the 3D momentum operator the null term ( Eq.(\ref{gamma defined})) split among the two components. Now, assume that a very strong potential $V(q_{2},q_3)$ confines the particle to the curve making the energy needed to excite the normal degrees of freedom very high, thus rendering the dynamics essentially 1-dimensional along the curve. Thus, take the limit $q_{2},q_3\rightarrow 0$ and at the same time drop the part of the kinetic energy operator that has the normal momentum operator, which is now Hermitian by construction, from the Hamiltonian. The resulting Hamiltonian -same as the one derived using the CJK approach - will be the Hermitian Hamiltonian on the curve with the geometric potential term that comes up being the piece that guarantees Hermicity. The origin of this geometric potential can be seen to stem from the ordering of the derivatives in the momentum operator in curvilinear coordinates. \\
The explicit form of the curve momentum operator that results when one pins the particle to the curve by takeing the limit $q_{2},q_3\rightarrow 0$ of the Hermitian 3D momentum operator parallel to the surface shows that it contains in addition to the derivative term a term proportional to the curvature of the curve and is - apparently- along the normal to the curve, Eq.(\ref{P curve Hermitian}), in complete analogy with the so called geometric momentum on a 2D surface embedded in 3D space. This additional term is seen to result, again, from ordering of the derivatives, and this curve geometrical momentum can be expressed as a symmetrical expression which is manifestly along the curve, Eq.(\ref{p curve symmetric}). This geometrical curve momentum is at the same time the kinematical momentum of the particle that comes out when one calculates the time derivative of the position operator (times the mass) using the Heisenberg equations of motion. The most general force operator of the particle is also found by taking the derivative of the geometrical curve momentum, and turns out to have  components along and normal to the curve. For a curve with constant curvature and torsion, this operator is shown to reduce to a -apparently- normal one, with a term that is a symmetrization of the classical centripetal force plus a quantum correction. All of the Hamiltonian, geometric curve momentum and force operators are calculated for the specific example of a spin zero particle confined to a cylindrical helix embedded in 3D space.\\
The above results are consistent with our previous results \cite{shikakhwa} for the construction of the Hamiltonian of a spin zero particle, both free and in an electromagnetic field, confined to an arbitrary 2D surface embedded in 3D Euclidean space applying the same approach, as well as the results  for the geometric momentum and force operators derived for free spin zero  particle confined to a 2D surface \cite{shikakhwa,shikakhwa symmetric}. The question of applying the current approach to the problem of spin one-half particle confined to a 2D surface embedded in 3D space was carried out in \cite{shikakhwa and chair1,shikakhwa and chair2} for orthogonal curvilinear coordinates; while that of confining such a particle to 1D curve embedded in 3D space is currently under  investigation.

\end{document}